# Reconfigurable chiral edge states in synthetic dimensions on an integrated photonic chip

Weiwei Liu,[1,§] Xiaolong Su,[1,§] Chijun Li,[1,§] Cheng Zeng,[1,§] Bing Wang,[1,*] Yongjie Wang,[1] Yufan Ding,[1] Chengzhi Qin,[1] Jinsong Xia,[1,†] and Peixiang Lu[1,2,‡]

[1]*Wuhan National Laboratory for Optoelectronics and School of Physics, Huazhong University of Science and Technology, Wuhan 430074, China*

[2]*Hubei Key Laboratory of Optical Information and Pattern Recognition, Wuhan Institute of Technology, Wuhan 430205, China*

**ABSTRACT**. Chiral edge states are a hallmark of topological physics, and the emergence of synthetic dimensions has provided ideal platforms for investigating chiral topology while overcoming the limitations of real space. Conventional studies have primarily concentrated on symmetric chiral behaviors, limited by complex and inflexible systems. Here we demonstrate a programmable integrated photonic platform to generate and manipulate reconfigurable chiral edge states in synthetic dimensions within a single lithium niobate microring resonator. Our system is realized by integrating independent frequency and pseudospin degrees of freedom in the dynamically modulated resonator, which features tunable artificial gauge potentials and long-range couplings. We demonstrate a variety of reconfigurable chiral behaviors in synthetic dimensions, including the realization and frustration of chiral edge states in a synthetic Hall ladder, the generation of imbalanced chiral edge currents, and the regulation of chiral behaviors among chirality, single-pseudospin enhancement, and complete suppression. This work paves the way for exploring chiral edge states in high-dimensional synthetic space on a programmable photonic chip, showing promising potential for applications in optical communications, quantum simulations, signal processing, and photonic neuromorphic computing.

## I. INTRODUCTION

Understanding chiral edge states is crucial for revealing spin-orbit coupling in topological physics, which has gained prominence in contexts such as topological insulators [1,2], valley-dependent optoelectronics [3,4] and topological photonics [5-7]. Experimental demonstrations indicate that modulating chiral edge states via external fields engenders exotic topological phenomena, including topological phase transitions [8,9], non-Abelian effects [10], chiral zero modes [11], and chiral Landau levels [12]. Specifically, nanophotonic systems have accelerated the study of chiral behaviors in photonic crystals [13], coupled resonators [14], metamaterials [15] and quasicrystals [16]. Despite the successful implementations, these approaches face inherent challenges due to inflexibility and complex fabrications, thus hindering manipulation of chiral edge states in nanophotonics.

Recently, photonic synthetic dimension has been a crucial innovation that leverages the degrees of freedom of photons to construct additional dimensions beyond real space, including frequency [17,18], arrival time of light pulses [19], mode [20] and orbital angular momentum [21]. This idea has garnered broad attention in condensed-matter physics [22,23], topological photonics [20] and non-Hermitian photonics [24,25]. In particular, the synthetic frequency dimension holds promise for simulating lattice dynamics [17] and bosonic transport [26], and the programmable lattices allow for performing reconfigurable photonic simulations in synthetic high-dimensional systems [7,27-30]. One of the significant breakthroughs has been implementation of controllable artificial gauge potentials to investigate chiral dynamics in a fiber loop, including spin-momentum locking, topological phase transition and chiral currents [9]. However, the absence of dynamic control over artificial gauge potentials and couplings constrains operational tunability, while limited integrability precludes deployment in high-speed regimes like optical communications, quantum information processing, and neuromorphic computing. Therefore, to establish a programmable integrated photonic platform for exploring and manipulating chiral edge states in synthetic dimensions remains challenging.

In this Letter, we demonstrate the generation and manipulation of reconfigurable chiral edge states on a programmable integrated photonic platform, created by coupling dual frequency lattices with opposite pseudospins in a thin-film lithium niobate (TFLN) microring resonator. Utilizing tunable artificial gauge potentials and couplings, we develop various strategies to manipulate chiral edge states on chip. First, we synthesize a chiral Hall ladder to investigate and frustrate the chiral edge currents with long-range couplings. We then create a synthetic misaligned interface and achieve imbalanced chiral edge currents with additional coupling on the Hall ladder. Finally, we construct a coupled Hall ladder by introducing next-nearest-neighbor couplings, showing that the chiral behaviors can be switched among chirality, single-pseudospin enhancement and complete suppression by controlling artificial gauge potentials.

*Contact author: wangbing@hust.edu.cn
†Contact author: jsxia@hust.edu.cn
‡Contact author: lupeixiang@hust.edu.cn
§These authors contributed equally to this work

## II. RESULTS

***Device design and theoretical model.*** Figure 1(a) depicts a microring cavity that supports a series of resonant frequency modes in both clockwise (CW) and counterclockwise (CCW) directions, corresponding to synthetic frequency and pseudospin dimensions respectively [9,31]. To couple individual frequency modes, reversed traveling-wave modulations are driven with frequency being an integer multiple of the cavity's free spectral range (FSR) [32]. Notably, the wavevector-matching principle permits independent modulation of opposite pseudospins [33,34], and coupling between the pseudospins is achieved by bringing two parts of the microring waveguide closer to each other [34]. Consequently, we realize a synthetic two-dimensional (2D) lattice resembling a series of reconfigurable frequency edges, as shown in Fig. 1(b).

To demonstrate reconfigurable frequency edges in synthetic dimensions, we fabricate the device on a TFLN platform (see Supplemental Material [34]), for advantageous electro-optic property. Recent advancements in integrated TFLN modulator have promoted modulation efficiency and bandwidth [35], generating considerable interest in applications including integrated microcomb [36], optical parametric oscillator [37], and microwave photonic engine [38]. Figure 1(c) presents a microscopic image of the fabricated device. Transmission spectrum of the passive cavity [Fig. 1(d)] reveals a quality factor of $3.5\times10^5$ and a FSR of ~9.34 GHz. The pseudospin coupling strength ($K$) is designed as 1/10 of the FSR, resulting in a resonant peak splitting of $2K$ = 1.82 GHz [Fig. 1(e)]. Under traveling-wave modulation, the system can be described with a tight-binding Hamiltonian [9,34]

$$H=-\sum_{m,s}\left[\omega_m a_{m,s}^{\dagger}a_{m,s}+\sum_{l=1}(J_s(t)a_{m,s}^{\dagger}a_{m-l,s}+\text{H.c.})\right], \qquad (1)$$
$$-K\sum_{m}(a_{m,\uparrow}^{\dagger}a_{m,\downarrow}+\text{H.c.})$$

where $a_{m,s}$ and $a_{m,s}^{\dagger}$ are the annihilation and creation operators for the $m$-order frequency mode of the cavity ($\omega_m$ = $m\Omega_R$), with pseudospin $s\in\{\uparrow(\text{CW}),\downarrow(\text{CCW})\}$. $\Omega_R/2\pi$ denotes FSR of the microring. $J_s(t)$ represent coupling strengths along frequency dimension for individual pseudospins, which are related to the traveling-wave modulations. For simplicity, the coupling strengths are assumed to be independent on mode index [31]. H.c. is the Hermitian conjugate. $K$ represents effective coupling strength between the pseudospin modes. To obtain a time-independent Hamiltonian, we transform to the interaction picture by defining $\tilde{a}_{m,s}=a_{m,s}e^{-im\Omega_R t}$. Then by applying the rotating-wave approximation and utilizing the Fourier transform $\tilde{a}_{k,s}=\sqrt{\frac{\Omega_R}{2\pi}}\sum_m e^{im\Omega_R k}\tilde{a}_{m,s}$, the Hamiltonian can be written in quasi-momentum space ($k$-space) as $H=\sum_k \tilde{a}_k^{\dagger}H_k\tilde{a}_k$, with [34]

$$H_k=-\begin{bmatrix} J_{\uparrow}(k) & K \\ K & J_{\downarrow}(k)\end{bmatrix}, \qquad (2)$$
$$\tilde{a}_k=[\tilde{a}_{k,\uparrow},\tilde{a}_{k,\downarrow}]^T$$

where $J_s(k)=\sum_l J_{s,l}\cos(l\Omega_R k+\varphi_{s,l})$, $J_{s,l}$ denotes the $l$-order frequency components of $J_s(t)$, and $\varphi_{s,l}$ is the corresponding initial phase. Actually, $H_k$ can be mapped onto a spin-orbit coupling (SOC) system by rewriting as $H_k=-\varepsilon(k)\cdot\mathbb{I}-B_{\text{soc}}\cdot\sigma$ [9], with $\varepsilon(k)=[J_{\uparrow}(k)+J_{\downarrow}(k)]/2$, $B_{\text{soc}}=\{K, 0, [J_{\uparrow}(k)-J_{\downarrow}(k)]/2\}$, $\mathbb{I}$ representing a 2 × 2 identity matrix, and $\sigma=[\sigma_x, \sigma_y, \sigma_z]$ denoting the Pauli matrix. Specifically, the eigenvalues of $H_k$ correspond to the energy band in $k$-space. Therefore, the synthetic frequency lattices and corresponding energy bands are governed by modulation signals. By controlling the modulations, reconfigurable edge states can be conveniently constructed and manipulated to investigate the chiral features.

***Synthesizing chiral Hall ladder in frequency and pseudospin dimensions.*** Hall ladder is one of the most important physical models for realizing topological chiral edge modes of a 2D quantum Hall insulator described by the Harper-Hofstadter Hamiltonian [8,9]. In our system, the Hall ladder is constructed by defining the modulation (coupling) as $J_{\uparrow}(t)=J\cos(\Omega_R t-\varphi_0/2)$, $J_{\downarrow}(t)=J\cos(\Omega_R t+\varphi_0/2)$ [Figs. 2(a) and 2(b)]. Specifically, the phase difference between modulation signals $\varphi_0$ will introduce an artificial gauge potential in the ladder [18].

Figure 2(c) shows energy band of the synthetic Hall ladder calculated using Equation (2), $J/K$ = 1.5, $\varphi_0$ = $0.75\pi$, with color shading quantifying the population of CW modes. Experimentally, the energy band can be directly obtained using time-resolved band structure spectroscopy [9,31,39]. A tunable continuous-wave laser is used to excite the CW pseudospin, and transmitted time-domain signal is measured to map out the band structure at drop port through continuously sweeping the frequency detuning $\Delta\omega$ [34]. The measured band structure [Fig. 2(d)] shows qualitative agreement with the theoretical result, and also with the numerical result [Fig. 2(e)] by time-domain coupled mode theory (TCMT) [9,31,40,41]

$$\begin{aligned}
\frac{\partial b_{m,\uparrow}}{\partial t} &= (im\Omega_R - \gamma/2)b_{m,\uparrow} + i\sum_n J_\uparrow(t)b_{n,\uparrow} + iKb_{m,\downarrow} + i\sqrt{\gamma_e}S_m^{(in)}e^{i\omega_m t} \\
\frac{\partial b_{m,\downarrow}}{\partial t} &= (im\Omega_R - \gamma/2)b_{m,\downarrow} + i\sum_n J_\downarrow(t)b_{n,\downarrow} + iKb_{m,\uparrow} \\
S^{(out)} &= i\sqrt{\gamma_e}\sum_m b_{m,\uparrow}
\end{aligned}, \quad (3)$$

where $b_{m,s}$ denotes the complex amplitudes in corresponding pseudospins with frequency $m\Omega_R$. $\gamma$ is the total loss of the cavity, and $\gamma_e$ represents external loss introduced by bus waveguide. $S_m^{(in)}$ is the input amplitude in CW direction with frequency $\omega_m = m\Omega_R + \Delta\omega$. $S^{(out)}$ is the complex amplitude of output light fields at drop port. As can be observed, in the lower band, the CW (CCW) modes predominantly occupy positive (negative) $k$-states in quasi-momentum space, manifesting spin-momentum locking. Under an effective magnetic flux, photons in two legs tend to propagate in opposite directions, resulting in chiral edge current. For quantitative characterization, chiral current is defined as $j_C = \sum_{m>m0}P(m) - \sum_{m<m0}P(m)$ [9], where $m_0$ represents index of the ring resonance closest to the input laser frequency. $P(m)$ is optical power at frequency $m\Omega_R$, extracted from measured optical spectra in Fig. 2(f). Measured chiral current $j_C$ and corresponding fitting using TCMT are presented in Fig. 2(g). In the lower (higher) energy band, the current for CW pseudospin is positive (negative), indicating that photons preferentially evolve to higher (lower) frequencies, which represents a hallmark of the photonic chiral edge current.

Specifically, the flexibility of our system permits independent modulation of the lattice constants for each pseudospin. As an extension, long-range coupling is introduced into one leg of the Hall ladder by defining the modulation as $J_\uparrow(t) = J\cos(\Omega_R t)$ and $J_\downarrow(t) = J\cos(N\Omega_R t + \varphi_0)$ ($N = 2, 3, 4, 5, \ldots, 10$), which creates a hetero-bilayer interface in synthetic dimension. In this scenario, the edge currents exhibit an overall decrease, indicating a frustration by long-range couplings (See Supplemental Materials for details [34]).

***Imbalanced chiral edge currents in synthetic misaligned interface.*** Chirality imbalance has become a compelling topic in topological physics, which has been realized through symmetry breaking in photonic metamaterials and fermionic systems [42,43]. Here, we demonstrate emergence of imbalanced chiral edge currents by imposing an additional coupling to one pseudospin of the system. Specifically, the modulation signals are adopted as $J_\uparrow(t) = J_1\cos(\Omega_R t + \varphi_1) + J_2\cos(2\Omega_R t + \varphi_2)$, $J_\downarrow(t) = J_2\cos(2\Omega_R t + \varphi_3)$ [Fig. 3(a)]. This configuration gives rise to a misaligned interface composed of two sub-Hall ladders and a triangular ladder [Fig. 3(b)], in which the breaking of symmetry between two pseudospins suggests a formation of imbalanced chiral characteristic.

To explore chiral features of the synthetic misaligned interface, an effective magnetic flux is applied to the sub-Hall ladders by defining $\varphi_1 = \varphi_2 = 0$, $\varphi_3 = -0.7\pi$. Figure 3(c) shows theoretically calculated energy band. Correspondingly, the measured and numerically calculated results are presented in Figs. 3(d)-3(e). The energy bands retain characteristic chiral features, as indicated by the dashed box in Fig. 3(c). However, introduction of the nearest-neighbor coupling leads to a deformation of the energy band. Consequently, the group velocity of CCW-dominant state becomes larger than that of CW-dominant state in the upper band, thereby supporting imbalanced chiral edge currents. To validate the prediction, we measure the optical spectra and chiral current [Figs. 3(f)-3(g)]. In the lower (upper) band, the chiral current for CW (CCW) mode is enhanced relative to CCW (CW) mode, indicating a net imbalance between opposite pseudospins. Furthermore, reversing direction of the effective magnetic flux inverts the current direction on both legs, while maintaining the imbalanced chiral characteristic [34].

***Manipulation of chiral edge states in coupled Hall ladders.*** Through introducing the next-nearest-neighbor coupling into Hall ladder, we construct a pseudo-three-dimensional lattice. Concretely, the modulation signals are defined as $J_{\uparrow,\downarrow}(t) = J_a\cos(\Omega_R t + \varphi_{\uparrow,\downarrow}) + J_b\cos(2\Omega_R t + \phi_{\uparrow,\downarrow})$. Correspondingly, this configuration results in two Hall ladders with lattice constants of $\Omega_R$ and $2\Omega_R$ respectively, along with a triangular ladder formed on each leg. These sub-ladders constitute a coupled Hall ladder, as illustrated in Fig. 4(a). Specifically, there are four effective magnetic fluxes, including those associated with the Hall ladders having lattice constant of $\Omega_R$ (denoted $\theta_1$) and $2\Omega_R$ (denoted $\theta_2$), as well as those corresponding to the triangular ladder in CCW (denoted $\theta_3$) and CW direction (denoted $\theta_4$). These fluxes are related to the modulation signals by $\theta_1 = \varphi_\downarrow - \varphi_\uparrow$, $\theta_2 = \phi_\uparrow - \phi_\downarrow$, $\theta_3 = \phi_\downarrow - 2\varphi_\downarrow$ and $\theta_4 = 2\varphi_\uparrow - \phi_\uparrow$. By configuring these magnetic fluxes, chiral edge states can be conveniently manipulated, enabling control over chirality, chiral triviality, single-pseudospin enhancement and complete suppression, which are summarized in a phase diagram [Fig. 4(b)], with $X = \mathrm{sgn}(\theta_2) - \mathrm{sgn}(\theta_1)$ and $Y = \mathrm{sgn}(\theta_3) - \mathrm{sgn}(\theta_4)$.

In the first case, we apply effective magnetic fluxes in the same direction in both Hall ladders ($\varphi_\uparrow = -\varphi_\downarrow = -0.15\pi$, $\phi_\uparrow = -\phi_\downarrow = -0.3\pi$). The coupling strengths are set as $J_a/K = 0.3$, $J_b/K = 0.21$, remaining unchanged for the following cases. Figure 4(c) indicates that the energy band is similar with that of the Hall ladder. Consequently, the edge currents exhibit chiral behavior, with opposite directions for the upper and lower bands in CW pseudospin modes [Figs. 4(d) and 4(e)]. Based on the first case, we introduce an additional effective magnetic flux into the triangular ladders ($\varphi_\uparrow = \phi_\downarrow = 0$, $\phi_\uparrow = -2\varphi_\downarrow = 0.3\pi$). In this case, the Hall ladders generate opposing chiral edge currents in each leg, while triangular ladders produce codirectional currents. Their interplay suppresses the current in one leg, enhancing chiral edge currents in the upper (lower) energy bands for CW (CCW) pseudospins. This single-pseudospin enhancement effect is demonstrated by the energy band and edge current in Figs. 4(f)-4(h). In the third case, effective magnetic fluxes of the Hall ladders are set in opposite directions ($\varphi_\uparrow = \phi_\downarrow = 0$, $\phi_\uparrow = 2\varphi_\downarrow = 0.3\pi$), which leads to cancellation and a complete suppression of the chiral edge currents, as shown in Figs. 4(i)-4(k). Additionally, for $|X| < |Y|$, effective magnetic fluxes in triangular ladders align, while in the Hall ladders cancel out. This results in nearly identical occupations for each pseudospin and dissolution of the edge states, corresponding to chiral triviality.

## III. DISCUSSION and CONCLUSIONS

We present an integrated photonic platform for simulating and manipulating chiral edge states in synthetic dimensions. While chiral edge states have been explored in other synthetic systems, such as fiber loops [9] and cold atoms [8], our platform offers several unique advantages. Firstly, such a platform enables the construction of chiral edge states through two independent physical dimensions within a single microring resonator, thereby simplifying system architecture. Secondly, leveraging traveling-wave modulations, we implement nonreciprocity to independently address pseudospin modes. This grants tunable effective gauge potentials and long-range couplings to manipulate the chiral edge states, even with complex modulation signals [34], surpassing the limitations of prior static nanophotonic systems. Thirdly, the reconfigurable chiral edge states on this platform will motivate further investigations of high-dimensional topological effects in nanophotonics, such as non-Hermitian chiral physics [24], Weyl semimetals [25], Floquet topological physics [44], and quantum correlations [45]. Therefore, this work will attract broad interdisciplinary interest from the community of condensed-matter physics, topological photonics, integrated photonics, and quantum physics.

In conclusion, we demonstrate reconfigurable chiral edge states in synthetic dimensions on an integrated TFLN photonic chip. These edge states are created by coupling frequency lattices with opposite pseudospins, and can be dynamically controlled through programmable artificial gauge potentials and couplings. As a result, we achieve a variety of chiral behaviors, including the realization and frustration of chiral edge states, the generation of imbalanced chiral currents, and the regulation of chiral behaviors among chirality, single-pseudospin enhancement, and complete suppression. This work integrates reconfigurable chiral edge states with monolithic nanophotonic architecture, independent synthetic dimensions, and programmable capability, which opens up new avenues for exploring topological physics in high-dimensional synthetic space on chip, and advocates potential applications in optical communications [46], quantum simulations [26,47], quantum signal processing [45], and photonic neuromorphic computing [48-51].

## ACKNOWLEDGMENTS

We thank Prof. Chi Zhang for the support in real-time oscilloscope measurements. The work is supported by the National Natural Science Foundation of China (No. 12374305, No. 62375097, No. 12021004). The authors gratefully acknowledged the Center of Micro-Fabrication and Characterization (CMFC) of Wuhan National Laboratory for Optoelectronics (WNLO) for their support in the nanofabrication of devices.

[1] B. A. Bernevig and S.-C. Zhang, Phys. Rev. Lett. **96**, 106802 (2006).

[2] S. Howard, L. Jiao, Z. Wang, N. Morali, R. Batabyal, P. Kumar-Nag, N. Avraham, H. Beidenkopf, P. Vir, E. Liu, C. Shekhar, C. Felser, T. Hughes, and V. Madhavan, Nat. Commun. **12**, 4269 (2021).

[3] L. Ju, Z. Shi, N. Nair, Y. Lv, C. Jin, J. Velasco, C. Ojeda-Aristizabal, H. A. Bechtel, M. C. Martin, A. Zettl, J. Analytis, and F. Wang, Nature **520**, 650 (2015).

[4] G.-J. Tang, X.-D. Chen, L. Sun, C.-H. Guo, M.-Y. Li, Z.-T. Tian, H.-H. Chen, H.-W. Wang, Q.-Y. Sun, Y.-D. Pan, X.-T. He, Y.-K. Su, and J.-W. Dong, Light: Sci. Appl. **13**, 166 (2024).

[5] A. B. Khanikaev and G. Shvets, Nat. Photonics **11**, 763 (2017).

[6] R. Barczyk, L. Kuipers, and E. Verhagen, Nat. Photonics **18**, 574 (2024).

[7] N. Parappurath, F. Alpeggiani, L. Kuipers, and E. Verhagen, Sci. Adv. **6**, aww4137 (2020).

[8] M. Atala, M. Aidelsburger, M. Lohse, J. T. Barreiro, B. Paredes, and I. Bloch, Nat. Phys. **10**, 588 (2014).

[9] A. Dutt, Q. Lin, L. Q. Yuan, M. Minkov, M. Xiao, and S. H. Fan, Science **367**, 59 (2020).

[10] Q. Liang, Z. Dong, J.-S. Pan, H. Wang, H. Li, Z. Yang, W. Yi, and B. Yan, Nat. Phys. **20**, 1738 (2024).

[11] H. Jia, R. Zhang, W. Gao, Q. Guo, B. Yang, J. Hu, Y. Bi, Y. Xiang, C. Liu, and S. Zhang, Science **363**, 148 (2019).

[12] H. Jia, M. Wang, S. Ma, R.-Y. Zhang, J. Hu, D. Wang, and C. T. Chan, Light: Sci. Appl. **12**, 165 (2023).

[13] A. B. Khanikaev, S. Hossein Mousavi, W.-K. Tse, M. Kargarian, A. H. MacDonald, and G. Shvets, Nat. Mater. **12**, 233 (2012).

[14] M. Hafezi, S. Mittal, J. Fan, A. Migdall, and J. M. Taylor, Nat. Photonics **7**, 1001 (2013).

[15] W. Zhao, Y. Zheng, C. Lu, Z. Wang, Y.-C. Liu, and S. Zhang, Phys. Rev. Lett. **133**, 233801 (2024).

[16] Y. Zhang, Z. Lan, L. Hu, Y. Shu, X. Yuan, P. Guo, X. Peng, W. Chen, and J. Li, Opt. Lett. **48**, 2229 (2023).

[17] L. Q. Yuan, Q. Lin, M. Xiao, and S. H. Fan, Optica **5**, 1396 (2018).
[18] C. Qin, F. Zhou, Y. Peng, D. Sounas, X. Zhu, B. Wang, J. Dong, X. Zhang, A. Alù, and P. Lu, Phys. Rev. Lett. **120**, 133901 (2018).
[19] X. Hu, S. Wang, C. Qin, C. Liu, L. Zhao, Y. Li, H. Ye, W. Liu, S. Longhi, P. Lu, and B. Wang, Adv. Photon. **6**, 046001 (2024).
[20] E. Lustig, S. Weimann, Y. Plotnik, Y. Lumer, M. A. Bandres, A. Szameit, and M. Segev, Nature **567**, 356 (2019).
[21] X.-W. Luo, X. Zhou, J.-S. Xu, C.-F. Li, G.-C. Guo, C. Zhang, and Z.-W. Zhou, Nat. Commun. **8**, 16097 (2017).
[22] T. Ozawa and H. M. Price, Nat. Rev. Phys. **1**, 349 (2019).
[23] L. Du, Y. Zhang, J.-H. Wu, A. F. Kockum, and Y. Li, Phys. Rev. Lett. **128**, 223602 (2022).
[24] K. W. A. D. S. Fan, Science **371**, 1240 (2021).
[25] W. Song, S. Wu, C. Chen, Y. Chen, C. Huang, L. Yuan, S. Zhu, and T. Li, Phys. Rev. Lett. **130**, 043803 (2023).
[26] A. Senanian, L. G. Wright, P. F. Wade, H. K. Doyle, and P. L. McMahon, Nat. Phys. **19**, 1333 (2023).
[27] J. Suh, G. Kim, H. Park, S. Fan, N. Park, and S. Yu, Phys. Rev. Lett. **132**, 033803 (2024).
[28] L. Yuan, Q. Lin, A. Zhang, M. Xiao, X. Chen, and S. Fan, Phys. Rev. Lett. **122**, 083903 (2019).
[29] D. Cheng, E. Lustig, K. Wang, and S. Fan, Light: Sci. Appl. **12**, 158 (2023).
[30] D. Cheng, K. Wang, C. Roques-Carmes, E. Lustig, O. Y. Long, H. Wang, and S. Fan, Nature **637**, 52 (2025).
[31] A. Dutt, M. Minkov, Q. Lin, L. Yuan, D. A. B. Miller, and S. Fan, Nat. Commun. **10**, 3122 (2019).
[32] L. Yuan, Y. Shi, and S. Fan, Opt. Lett. **41**, 741 (2016).
[33] M. Yu, R. Cheng, C. Reimer, L. He, K. Luke, E. Puma, L. Shao, A. Shams-Ansari, X. Ren, H. R. Grant, L. Johansson, M. Zhang, and M. Lončar, Nat. Photonics **17**, 666 (2023).
[34] See Supplemental Material at http://xxxxxx for (I) the theory model, (II) methods and experimental details, (III) frustration of edge current through long-range coupling in Hall ladder, (IV) reversal of imbalanced chiral edge states by opposite effective magnetic flux, (V) energy bands and chiral edge states under triangular-wave modulation, (VI) evolution of Bloch states in synthetic chiral edges, (VII) error analysis of chiral edge current results, which includes Refs. [52-54].
[35] A. Boes, L. Chang, C. Langrock, M. Yu, M. Zhang, Q. Lin, M. Lončar, M. Fejer, J. Bowers, and A. Mitchell, Science **379**, eabj4396 (2023).
[36] M. Zhang, B. Buscaino, C. Wang, A. Shams-Ansari, C. Reimer, R. Zhu, J. M. Kahn, and M. Lončar, Nature **568**, 373 (2019).
[37] H. S. Stokowski, D. J. Dean, A. Y. Hwang, T. Park, O. T. Celik, T. P. McKenna, M. Jankowski, C. Langrock, V. Ansari, M. M. Fejer, and A. H. Safavi-Naeini, Nature **627**, 95 (2024).
[38] H. Feng, T. Ge, X. Guo, B. Wang, Y. Zhang, Z. Chen, S. Zhu, K. Zhang, W. Sun, C. Huang, Y. Yuan, and C. Wang, Nature **627**, 80 (2024).
[39] Z.-A. Wang, Y.-T. Wang, X.-D. Zeng, J.-M. Ren, W. Liu, X.-H. Wei, Z.-P. Li, Y.-Z. Yang, N.-J. Guo, L.-K. Xie, J.-Y. Liu, Y.-H. Ma, J.-S. Tang, Z.-W. Zhou, C.-F. Li, and G.-C. Guo, Phys. Rev. Lett. **133**, 233805 (2024).
[40] S. Fan, W. Suh, and J. D. Joannopoulos, J. Opt. Soc. Am. A **20**, 569 (2003).
[41] Z. Zhao, C. Guo, and S. Fan, Phys. Rev. A **99**, 033839 (2019).
[42] Q. Li, D. E. Kharzeev, C. Zhang, Y. Huang, I. Pletikosić, A. V. Fedorov, R. D. Zhong, J. A. Schneeloch, G. D. Gu, and T. Valla, Nat. Phys. **12**, 550 (2016).
[43] X. Zhou, S. Li, C. Hu, G. Wang, and B. Hou, Appl. Phys. Lett. **123**, 111702 (2023).
[44] S. K. Sridhar, S. Ghosh, D. Srinivasan, A. R. Miller, and A. Dutt, Nat. Phys. **20**, 843 (2024).
[45] U. A. Javid, R. Lopez-Rios, J. Ling, A. Graf, J. Staffa, and Q. Lin, Nat. Photonics **17**, 883 (2023).
[46] M. He, M. Xu, Y. Ren, J. Jian, Z. Ruan, Y. Xu, S. Gao, S. Sun, X. Wen, L. Zhou, L. Liu, C. Guo, H. Chen, S. Yu, L. Liu, and X. Cai, Nat. Photonics **13**, 359 (2019).
[47] T. Dai, A. Ma, J. Mao, Y. Ao, X. Jia, Y. Zheng, C. Zhai, Y. Yang, Z. Li, B. Tang, J. Luo, B. Zhang, X. Hu, Q. Gong, and J. Wang, Nat. Mater. **23**, 874 (2024).
[48] J. Feldmann, N. Youngblood, M. Karpov, H. Gehring, X. Li, M. Stappers, M. Le Gallo, X. Fu, A. Lukashchuk, A. S. Raja, J. Liu, C. D. Wright, A. Sebastian, T. J. Kippenberg, W. H. P. Pernice, and H. Bhaskaran, Nature **589**, 52 (2021).
[49] X. Y. Xu, M. X. Tan, B. Corcoran, J. Y. Wu, A. Boes, T. G. Nguyen, S. T. Chu, B. E. Little, D. G. Hicks, R. Morandotti, A. Mitchell, and D. J. Moss, Nature **589**, 44 (2021).
[50] A. Sludds, S. Bandyopadhyay, Z. J. Chen, Z. Z. Zhong, J. Cochrane, L. Bernstein, D. Bunandar, P. Ben Dixon, S. A. Hamilton, M. Streshinsky, A. Novack, T. Baehr-Jones, M. Hochberg, M. Ghobadi, R. Hamerly, and D. Englund, Science **378**, 270 (2022).
[51] S. Bandyopadhyay, A. Sludds, S. Krastanov, R. Hamerly, N. Harris, D. Bunandar, M. Streshinsky, M. Hochberg, and D. Englund, Nat. Photonics **18**, 1335 (2024).
[52] R. Ye, Y. Y. He, G. Z. Li, L. J. Wang, X. X. Wu, X. Qiao, Y. L. Zheng, L. Jin, D. W. Wang, L. Q. Yuan, and X. F. Chen, Light: Sci. Appl. **14**, 39 (2025).
[53] D. Hügel and B. Paredes, Phys. Rev. A **89**, 023619 (2014).
[54] M. Mancini, G. Pagano, G. Cappellini, L. Livi, M. Rider, J. Catani, C. Sias, P. Zoller, M. Inguscio, M. Dalmonte, and L. Fallani, Science **349**, 1510 (2015).

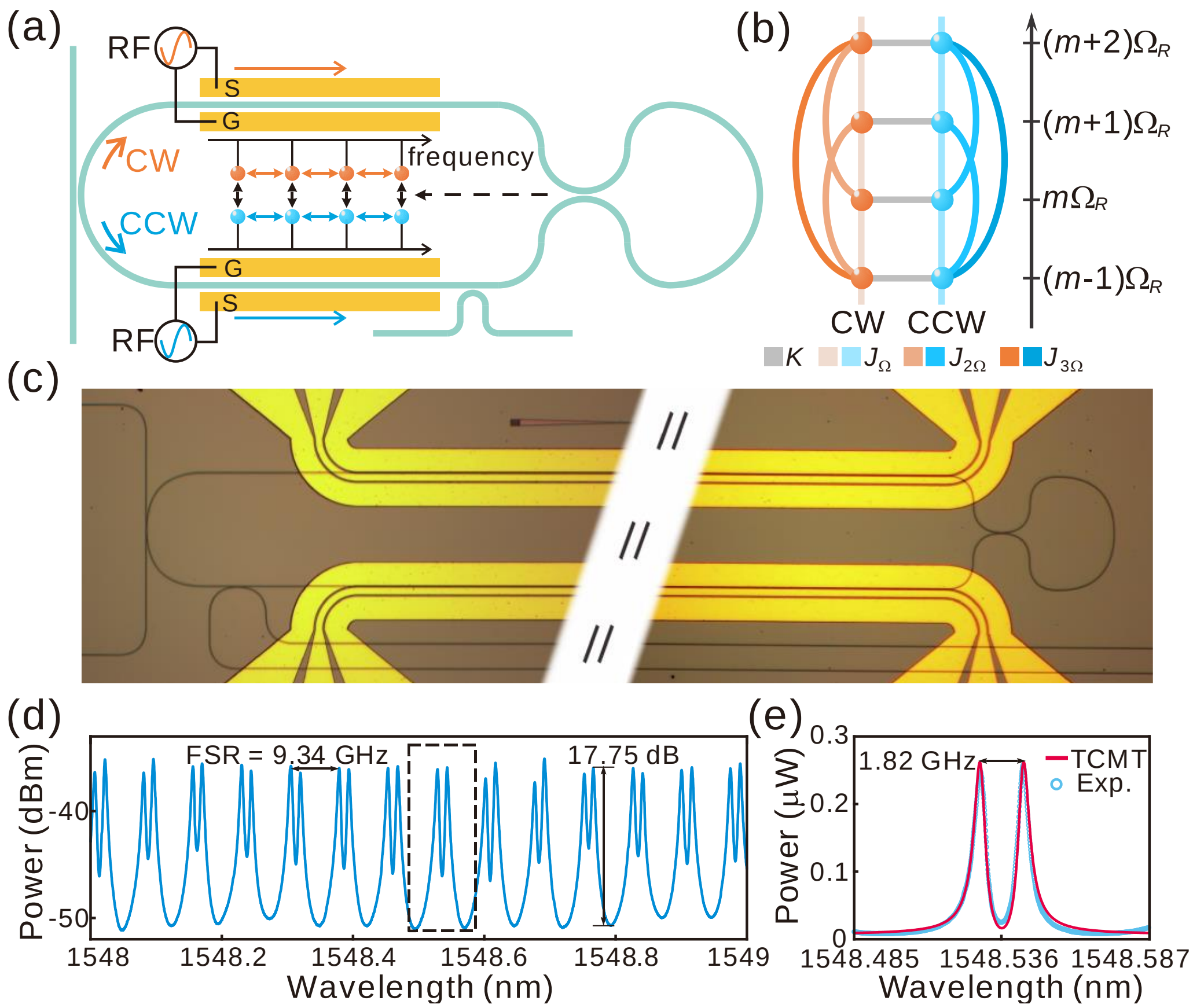

**Fig. 1. Concept of reconfigurable chiral edge states via synthetic dimensions in an integrated photonic platform.** (a) Structural schematic for realizing synthetic frequency and pseudospin dimensions in a coupled microring resonator. (b) Diagram for creating reconfigurable chiral edge states. (c) Microscopic image of a device fabricated on TFLN platform. (d) Measured transmission spectrum of the passive cavity. (e) An enlarged image of the transmission spectrum.

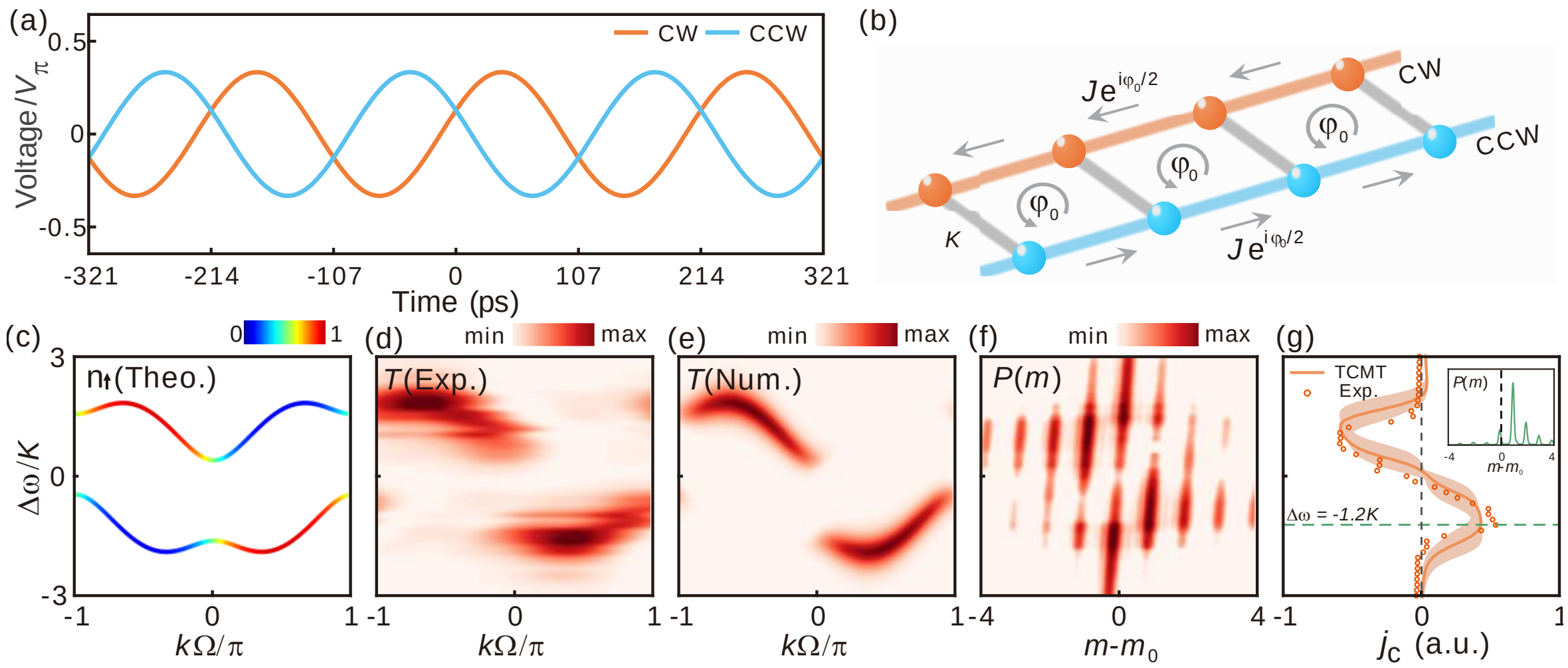

**Fig. 2. Chiral edge states in a synthetic Hall ladder.** (a) Plots of modulation signals. (b) Illustration of a synthetic Hall ladder. (c) Theoretical energy band, $J/K = 1.5$, $\varphi_0 = 0.75\pi$. (d) Experimentally measured and (e) numerically calculated time-resolved band structures. (f) Experimentally measured optical spectra mapped with laser detuning ($\Delta\omega$). (g) Chiral edge current extracted from measured optical spectra, and corresponding numerical fitting with root mean square error (RMSE) of 0.14. The shadow region represents simulated deviations, and the inset represents a typical optical spectrum at $\Delta\omega = -1.2K$.

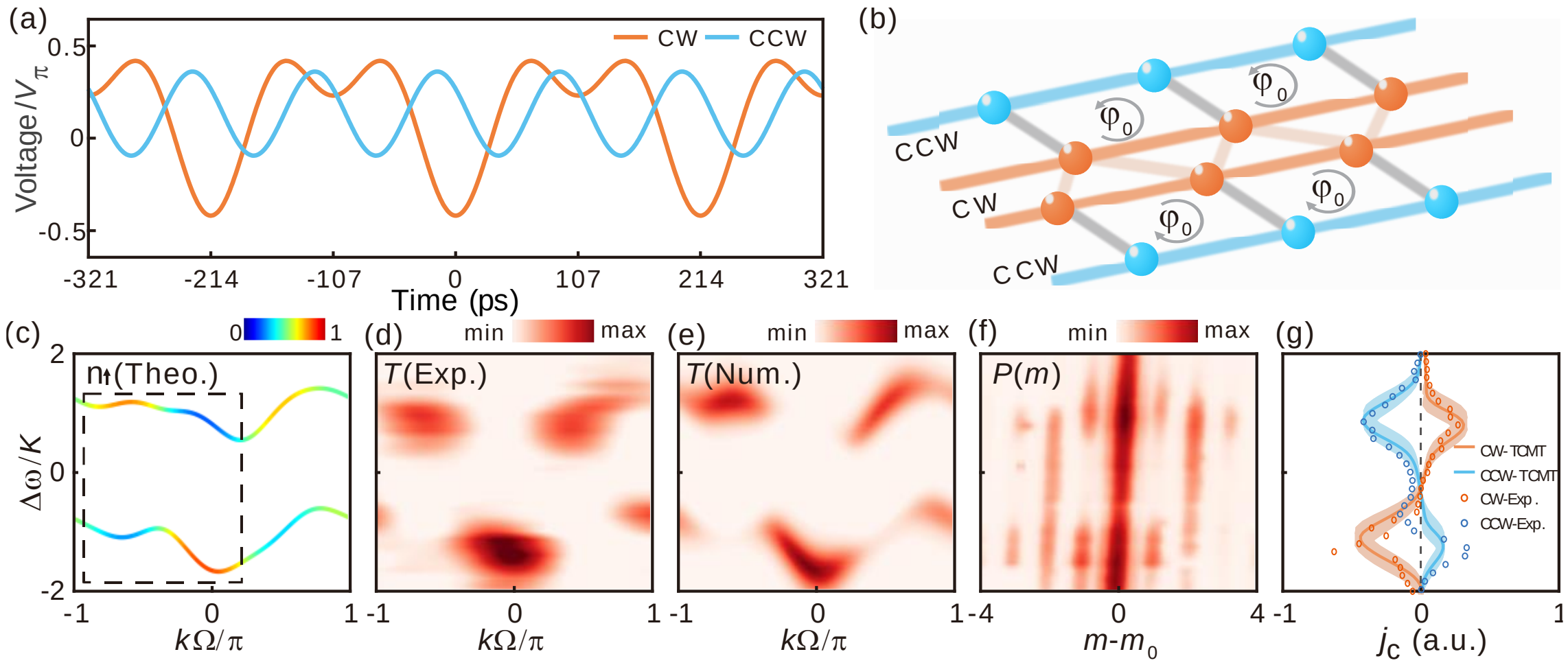

**Fig. 3. Observation of imbalanced chiral edge currents.** (a) Plots of modulation signals. (b) Illustration of a synthetic misaligned interface. (c) Theoretical energy band, $J_1/K$ = 0.65, $J_2/K$ = 0.46, $\varphi_1 = \varphi_2 = 0$, $\varphi_3 = -0.7\pi$. (d) Experimentally measured and (e) numerically calculated time-resolved band structures. (f) Experimentally measured optical spectra mapped with laser detuning ($\Delta\omega$). (g) Chiral edge current extracted from measured optical spectra, and corresponding numerical fittings with RMSE of 0.08. The shadow regions represent simulated deviations.

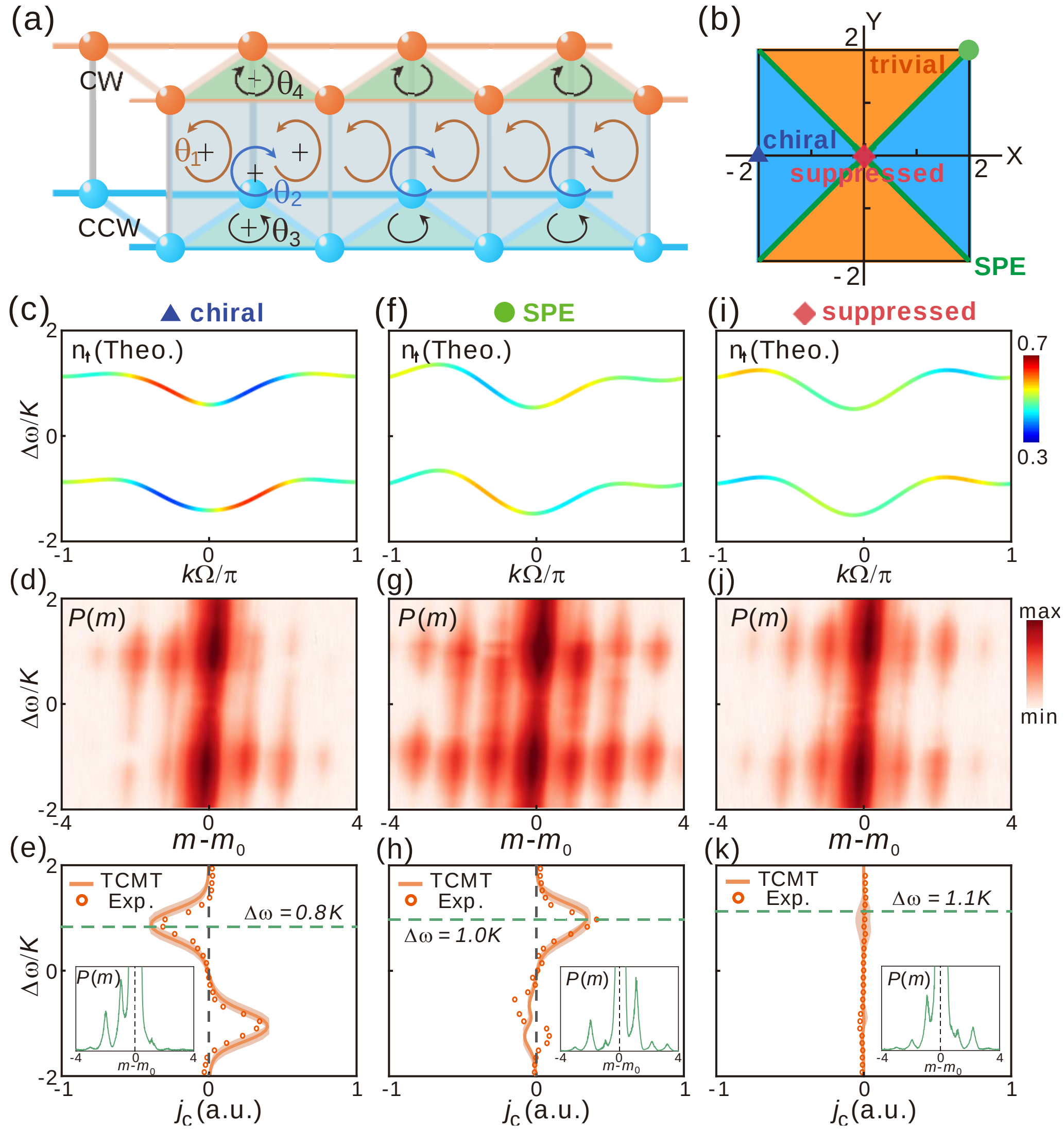

**Fig. 4. Manipulation of chiral edge states in coupled Hall ladders.** (a) Illustration of coupled Hall ladders. (b) Phase diagram of the chiral edge current. Theoretical energy bands, measured spectra mappings and chiral edge currents for (c-e) chirality ($\varphi_\uparrow = -\varphi_\downarrow = -0.15\pi$, $\phi_\uparrow = -\phi_\downarrow = -0.3\pi$), (f-h) single-pseudospin enhancement (SPE, $\varphi_\uparrow = \phi_\downarrow = 0$, $\phi_\uparrow = -2\varphi_\downarrow = 0.3\pi$), and (i-k) complete suppression ($\varphi_\uparrow = \phi_\downarrow = 0$, $\phi_\uparrow = 2\varphi_\downarrow = 0.3\pi$). RMSEs between numerical and experimental results are 0.07, 0.07, and 0.01 for (e), (h) and (k) respectively. The shadow regions represent simulated deviations. The insets in (e), (h) and (k) represent typical optical spectra measured at $\Delta\omega = 0.8K$, $1.0K$, and $1.1K$, respectively.